\documentclass{eas}

\usepackage{graphicx}
\usepackage{natbib}
\usepackage{txfonts}
\begin{document}

\title{Exoplanets: Gaia and the importance of ground based spectroscopy follow-up}
\runningtitle{Benamati \etal: Exoplanets: Gaia and \dots}
\author{L. Benamati}
\address{Instituto de Astrof\'{\i}sica e Ci\^encia do Espa\c{c}o, Universidade do Porto, Rua das Estrelas, 4150-762 Porto, Portugal; \email{Lisa.Benamati@astro.up.pt}}
\secondaddress{Departamento de F\'{\i}sica e Astronomia, Faculdade de Ci\^encias, Universidade do Porto, Portugal}
\author{V. Zh. Adibekyan}
\sameaddress{1}
\author{N.C. Santos}
\sameaddress{1,2}
\author{A. Sozzetti}
\address{INAF- Osservatorio Astrofisico di Torino, 10025 Pino Torinese, Italy}
\begin{abstract}
The search for extrasolar planets has developed rapidly and, today, more than 1700 planets have been found orbiting stars. Thanks to Gaia, we will collect high-accuracy astrometric orbits of thousands of new low-mass celestial objects, such as extra-solar planets and brown dwarfs. 
These measurements in combination with spectroscopy and  with present day and future extrasolar planet search programs (like HARPS,  ESPRESSO) will have a crucial contribution to several aspects of planetary astrophysics (formation theories, dynamical evolution, etc.). 
Moreover, Gaia will have a strong contribution on the stellar chemical and kinematic characterisation studies.
In this paper we present a short overview of the importance of Gaia in the context of exoplanet research. As preparatory work for Gaia, we will then present a study where we derived stellar parameters for a sample of field giant stars.
\end{abstract}

\maketitle

\section{Introduction}
The combination of Gaia astrometric data with the spectroscopic follow-up will allow us to improve our knowledge on exoplanets: astrometry and spectroscopy are complementary and the synergy of these two methods is really useful to study exoplanets (e.g.\citet{soz09_b}, \citet{soz10_b}). 
Because of the nature of the spectroscopic (radial velocity) method, only a lower limit on the planet mass is known from specroscopic measurements alone. Instead, the astrometry determines the inclination of the orbital plane with respect to the plane of the sky which is unresolved in radial velocity.
For this reason, combining radial velocity with astrometry we could calculate the mass of the planet and hence with Gaia we expect to improve the current landscape by adding thousands of new planets (\citet{latta02}).\\ 
But not only!
Astrometry searches for long$-$period planets around nearby stars of all spectral types. The spectral type, metallicity, and the rotational velocity of the star do not strongly condition the astromeric precision (e.g. \citet{mart10}, \citet{dum11}). 
In addition, astrometry can detect giant planets around active young stars, while for the RV, in principle, is more difficult due to the noise in the spectrum (\citet{des07}). 
Furthermore, the sensitivity of astrometry and RV to planetary signals is highly complementary.
The first one is more sensitive to planets far from the star (signal scales linearly with the semimajor axis $a$, while the second one is more sensitive to planets close to the star signal scales with $1/\sqrt(a)$). 
On the contrary, the radial velocity method is more sensitive than astrometry in detecting low-mass planets, particularly thanks to the new generation of high-precision instruments, such as HARPS and HARPS-N (\citet{pepe08}, \citet{cos12}). 
In preparation for Gaia, a recent study gauged the combinatory power of Hipparcos
astrometry and precision RVs on a sample of metal-poor binaries (\citet{ben13})\\
Spectroscopy is a powerful tool which can be used not only to detect a planet (via RV measurements), but also to characterize the host stars. 
In addition to a precise characterization of the planets, it is very important to collect information on their parent stars and having new several thousand of giant planets detections, thanks to Gaia, and knowing well the properties of the parent stars from the spectroscopy and astrometry, we will be able to investigate their characteristics as a function of stellar mass, metallicity, age, etc.. 
It is for instance known that the stellar properties influence both the frequency and properties of the known planets, a fact that has important consequences for the models of planet formation and evolution (\citet{san04}, \citet{adi13}, \citet{daw13}).
Note that here the role of Gaia will be huge too. For instance, Gaia parallaxes will allow to calculate precise surface gravities for the stars (e.g. \citet{soz07}, \citet{mor13}). \\

\section{Stellar parameters for a sample of giant stars: preparing for Gaia}
\begin{figure*}
\begin{center}
\begin{tabular}{c}
\includegraphics[angle=0,width=0.6\linewidth]{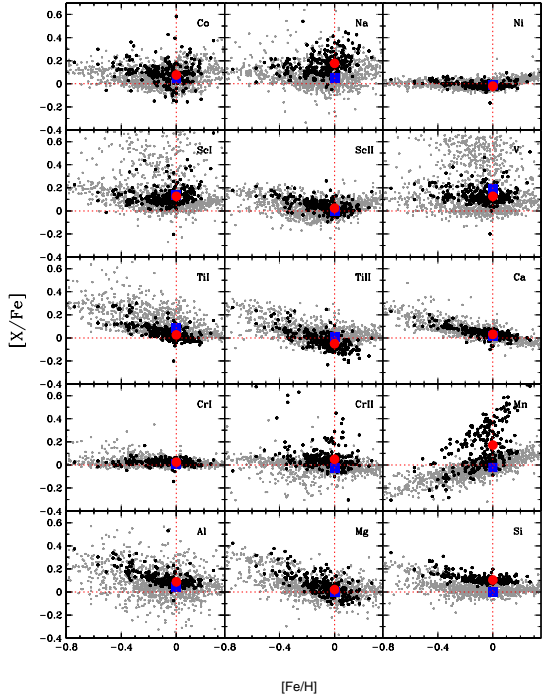}
\end{tabular}
\end{center}
\vspace{-1.cm}
\caption{[X/Fe] vs. [Fe/H] plots. The black dots represent the stars of the sample and the gray small dots represent stars from \citet{adi12} with \emph{$T{}_{\mathrm{eff}}$} = \emph{T$_{\odot}$$\pm$$500$ K}. The red circle and blue square show the average [X/Fe] value of stars with [Fe/H] = 0.0$\pm$0.1 dex. Each element is identified in the \emph{upper right corner} of the respective plot.}
\label{fig:c4_14}
\end{figure*}

Waiting for Gaia data, we studied giant stars screened for planets in RV surveys in the context of the CORALIE (\citet{udr00}) extrasolar planet search program analyzing the stellar parameters (effective temperature, microturbolence, surface gravity and metallicity) and also chemical abundances for 12 elements (Na, Mg, Al, Si, Ca, Ti, Cr, Ni, Co, Sc, Mn and V) in their atmospheres.
Since the sample is part of a planet search program using radial velocities, the goal of this study is to explore the star-planet connection (frequency, properties) for planets orbiting giant stars. In the meanwhile, the sample allows us to study the chemical evolution of the Galaxy.
The stars in the sample have effective temperatures 4700 $\lesssim$ Teff $\lesssim$ 5600 K, surface gravities 2.2 $\lesssim$ log$g$ $\lesssim$ 3.7 dex, microturbulence 1 $\lesssim$ $\xi_{t}$ $\lesssim$ 3.2 km s$^{-1}$ and they lie in the metallicity range of $-0.75$ $\lesssim$ [Fe/H] $\lesssim$ 0.3 dex (\citet{san15}, \citet{adi15}).\\
In Fig. \ref{fig:c4_14} we present the dependence of [X/Fe] on metallicity for our sample of giant stars and for FGK dwarf stars from \citet{adi12} to study the Galactic chemical evolution.
As one can see, for all the elements the general behavior of [X/Fe] with the metallicity is similar for giant and dwarf stars, and reflects the Galactic chemical evolution in the solar neighborhood. 
The largest offset is seen for Na and Mn, and a bit less in Si and Al. The Mn abundance was obtained by using only one line and it should be considered with caution. Moreover, the scatter in [Mn/Fe] is very high, indicating unrealistic abundances for some fraction of the stars.\\
To separate Galactic stellar populations we applied both a kinematic and chemical approach. 
For the kinematic approach , the selection of the thin disk, thick disk and halo stars was found using the method described in \citet{be03} and \citet{ro03}. We found that most of our stars belong to the thin disk (96$\%$ and 97$\%$).
For the chemical approach we use the abundances of $\alpha$-element since the thin and thick disk stars are also different in their $\alpha$-element content at a given metallicity. The [$\alpha$/Fe] versus [Fe/H] plot (not shown) for the sample stars confirms that most of our stars belong to the thin disk.

\section{Conclusion}
In summary, we found that for all the elements studied Galactic chemical evolution trends are similar for giant and dwarf stars, while for some species [X/Fe] values are shifted towards higher values at a fixed metallicity. Our analyis confirms the overabundance of Na (also Al and Si to a lesser degree) in giant stars compared to the field FGK dwarf stars from \citet{adi12}.
Our sample is manly composed by stars belonging to the thin disk and the analyzed sample includes still only one star with detected planetary companion. 
Before a significant number of planets are detected, this sample can be used as a homogeneous comparison sample to study planet occurrence around giant stars.

\end{document}